\documentclass[12pt]{article}
\usepackage{graphicx,color}
\graphicspath{{./Figs/}}

\begin{document}
\baselineskip=5.5mm
\newcommand{\be} {\begin{equation}}
\newcommand{\ee} {\end{equation}}
\newcommand{\Be} {\begin{eqnarray}}
\newcommand{\Ee} {\end{eqnarray}}
\renewcommand{\thefootnote}{\fnsymbol{footnote}}
\def\a{\alpha}
\def\b{\beta}
\def\g{\gamma}
\def\G{\Gamma}
\def\d{\delta}
\def\D{\Delta}
\def\e{\epsilon}
\def\k{\kappa}
\def\l{\lambda}
\def\L{\Lambda}
\def\t{\tau}
\def\om{\omega}
\def\Om{\Omega}
\def\s{\sigma}
\def\lg{\langle}
\def\rg{\rangle}
\parindent 0cm
\newcommand{\tblue}[1]{\textcolor{blue}{#1}}
\begin{center}
{\large {\bf Nonlinear response functions in an exponential trap model} }\\
\vspace{0.5cm}
\noindent
{\bf Gregor Diezemann} \\
email: diezeman@uni-mainz.de\\
{\it
Institut f\"ur Physikalische Chemie, Universit\"at Mainz,
Duesbergweg 10-14,\\ 55128 Mainz, FRG\\

}
\end{center}

\noindent
{\it
The nonlinear response to an oscillating field is calculated for a kinetic trap model with an exponential density of states and the results are compared to those for the model with a Gaussian density of states.
The calculations are limited to the high temperature phase of the model.
It is found that the results are qualitatively different only in a temperature range near the glass transition temperature $T_0$ of the exponential model.
While for the Gaussian model the choice of the dynamical variable that couples to the field has no impact on the shape of the linear response, this is different for the exponential model.
Here, it is found that also the relaxation time strongly depends on the variable chosen.
Furthermore, the modulus of the frequency dependent third-order response shows either a peak or exhibits a monotonuous decay from a finite low-frequency limit to a vanishing response at high frequencies depending on the dynamical variable.
For variables that give rise to a peak in the modulus it is found that its height either increases or decreases as a function of temperature, again depending on the details of the choice of the variable.
The peak value of the modulus shows a scaling behavior near $T_0$.
It is found that for some variables the low-frequency limit of the cubic response diverges at the glass transition temperature and also at a further temperature determined by the particular variable.
A recently proposed approximation that relates the cubic response to a four-time correlation function does not give reliable results due to a wrong estimate of the low-frequency limit of the response.
}

\vspace{1.5cm}

\section*{I. Introduction}
There have been many attempts to understand the heterogeneous dynamics in supercooled liquids and 
glasses, cf. the reviews\cite{Berthier:2011p6852,Sillescu99} and references therein.
In particular, the development of experimental techniques to probe higher-order time correlation functions have played a pivotal role in the development of our understanding of the nature of the heterogeneities\cite{SRS91, HWZS95, G13, SBLC96, G16, G23} and also the study of computer models allowed to gain insight into the structural and dynamical properties of these correlation functions\cite{Schroder03, Reichman07}.
Also a length scale associated with the heterogeneities has been extracted from specially designed NMR 
experiments\cite{Tracht98,Reinsberg01}.

In addition to these approaches an additional way to extract a length scale from a special four-point correlation function $\chi_4(t)$  has been introduced and discussed in detail\cite{Berthier05, Toninelli05, Berthier07a, Berthier07b}.

By relating the nonlinear (cubic) response $\chi_3(\om,T)$ to a four-point correlation function, Bouchaud and Biroli showed how to extract a length scale or equivalently the number of correlated particles, $N_{\rm corr}$, from measured nonlinear response 
functions\cite{Bouchaud05}.
The modulus of the cubic response function, $|\chi_3(\om,T)|$, was found to exhibits a hump-like structure which is assumed to be a distinctive feature of glassy correlations\cite{CrausteThibierge10,Brun11}.
It is found that the maximum of $|\chi_3(\om,T)|$ decreases with increasing temperature and it is assumed to be proportional to 
$N_{\rm corr}$.
If glassy correlations are absent, 'trivial' behavior is expected, i.e. a smooth decay of $|\chi_3(\om,T)|$ as a function of frequency.

It should be mentioned that nonlinear dielectric experiments on supercooled liquids have also been interpreted in a slightly different way with a stronger emphasis on the heterogeneous nature of the dynamics\cite{Richert06, Weinstein07,Wang07}.

A nonlinear response theory for Markov processes has been presented in ref.\cite{G75}, to be denoted as I in the following.
The theory was applied to the model of dipole reorientations in an asymmetric double well potential (ADWP-model)\cite{Frohlich49, G46}. 
For this model, $|\chi_3(\om,T)|$ exhibits trivial behavior except for a small temperature range in the vicinity of vanishing low-frequency limit $\chi_3(0,T)$ for finite asymmetry.
In addition, model calculations were presented for the well-studied trap model with a Gaussian density of states\cite{Dyre95, MB96, Denny03, G64, G71, G73} showing both, a peak or trivial behavior, depending on the variable chosen and on temperature. 
Furthermore, for some specific choice of the dynamical variable used to probe the dynamics, the peak-maximum increases as a function of temperature and for other choices it decreases.
The results of the model calculations suggest that a direct relation between the cubic response function and some type of glassy correlations cannot be shown to exist in these mean-field models.
Also other calculations employing specific models show a similar behavior, i.e. either the existence of a hump or a trivial 
decay\cite{Brun11b,Dejardin:2014}.

In addition, in ref.\cite{G76}, denoted as II in what follows, I considered various four-time correlation functions and a particular approximation for the cubic response for the Gaussian trap model.
According to the approximations employed by Bouchaud and Biroli\cite{Bouchaud05}, the most dominant contribution to the cubic response in the vicinity of a phase transition is related to a four-time correlation function.
For the Gaussian trap model, it was found in II that the corresponding relation does not give a sound description of $|\chi_3(\om,T)|$ due to a wrong estimate of the low-frequency behavior.

In the present paper, I will calculate the nonlinear response for the trap model with an exponential density of states (DOS) instead of a Gaussian DOS.
The most prominent difference between the two models is that the exponential DOS gives rise to a glass transition at a temperature $T_0$ below which the system cannot equilibrate.
Thus, there is a critical point in this model and one can investigate the nonlinear response in the vicinity of this point.
In the present paper, I will only consider the high temperature phase where equilibrium is always reached.
In the next Section, I will briefly recall the properties of the model and discuss the modifications in the linear response resulting from a specific choice of the dynamic variables.
Section III is devoted to a discussion of the nonlinear response and the paper closes with some conclusions.
\section*{II. Trap model with an exponential density of states}
The stochastic dynamics for the trap model is defined by the master equation (ME) for the conditional probability to find the system in the trap characterized by the trap energy $\e$ at time $t$ provided it was in trap $\e_0$ at $t_0$, $G(\e,t+t_0|\e_0,t_0)=G(\e,t|\e_0,0)\equiv G(\e,t|\e_0)$:
\be\label{ME.G}
{\dot G}(\e,t|\e_0)= -\k(\e)G(\e,t|\e_0)+\rho(\e)\!\int\!d\e'\k(\e')G(\e',t|\e_0) 
\ee
with the escape rate given by ($\k_\infty$ denotes an attempt frequency)
\be\label{k.T}
\k(\e)=\k_\infty e^{-\b\e}
\ee
The model with an exponential density of states (DOS) is defined by\cite{MB96}
\be\label{DOS.exp}
\rho(\e)\!=\!\b xe^{-\b x\e}
\quad\mbox{with}\quad 
x=T/T_0
\ee
where $\b=1/T$ and the Boltzmann constant is set to unity.
Here, $T_0$ denotes the characteristic temperature of the model, below which the equilibrium distribution
$p^{eq}(\e)=\lim_{t\to\infty}G(\e,t|\e_0)$ cannot be normalized (i.e. the integral $\int\!d\e\rho(\e)e^{\b \e}$ diverges for $x<1$).
Above $T_0$, one has
\be\label{peq}
p^{eq}(\e)=\b(x-1)e^{-\b(x-1)\e}
\ee
and below $T_0$ no equilibrium is reached and the system ages for all times.
The model exhibits a number of features that are reminiscent of what is observed in glassy systems\cite{MB96}.
In the present paper, I will solely consider temperatures above $T_0$, i.e. $x>1$ and aging is unimportant.

The two-time correlation function (2t-CF) of a variable $M(t)$ in general is given by:
\be\label{C2.Pi}
\lg M(t)M(t_0)\rg
=\int\!d\e\int\!d\e_0 M(\e)M(\e_0)G(\e,t-t_0|\e_0)p^{\rm eq}(\e_0)
\ee
The quantity $M(t)$ might for example represent a magnetization or a dipole moment.
In a naive picture one could for instance assume that high energy regions correspond to low density regions and that the dipole moment varies with the latter.
As in paper I and II \cite{G75, G76}, a Gausian approximation for the correlations of the dynamical variables $M(\e)$ will be used,
\be\label{MkMl.mit}
\lg M(\e)\rg=0
\quad\mbox{and}\quad
\lg M(\e)M(\e_0)\rg=\d(\e-\e_0)\lg M(\e)^2\rg
\ee
For a fully connected trap model one has (changing to the common notation\cite{MB96})
\be\label{Pi.t}
\Pi(t)=\int\!d\e\lg M(\e)^2\rg p^{\rm eq}(\e)e^{-\k(\e)t}
\ee
This function has a simple interpretation.
Each transition out of the trap with energy $\e$ completely decorrelates the variable and gives rise to a decay.
For $\lg M(\e)^2\rg=1$, it has been shown that the long-time behavior of $\Pi(t)$ is given by 
$\Pi(t)\sim t^{-(x-1)}$\cite{MB96}.

Throughout the present paper, I will use an Arrhenius-like energy dependence of $\lg M(\e)^2\rg$, that first has been considered by Fielding and Sollich\cite{FS02} in their treatment of the violations of the fluctuation disspiation theorem for the trap model and that I have used also in papers I and II\cite{G75, G76}:
\be\label{M2e.exp}
\lg M(\e)^2\rg=e^{-n\b\e}
\ee
where $n$ is an arbitrary real constant.

\begin{figure}[h!]
\centering
\includegraphics[width=8.0cm]{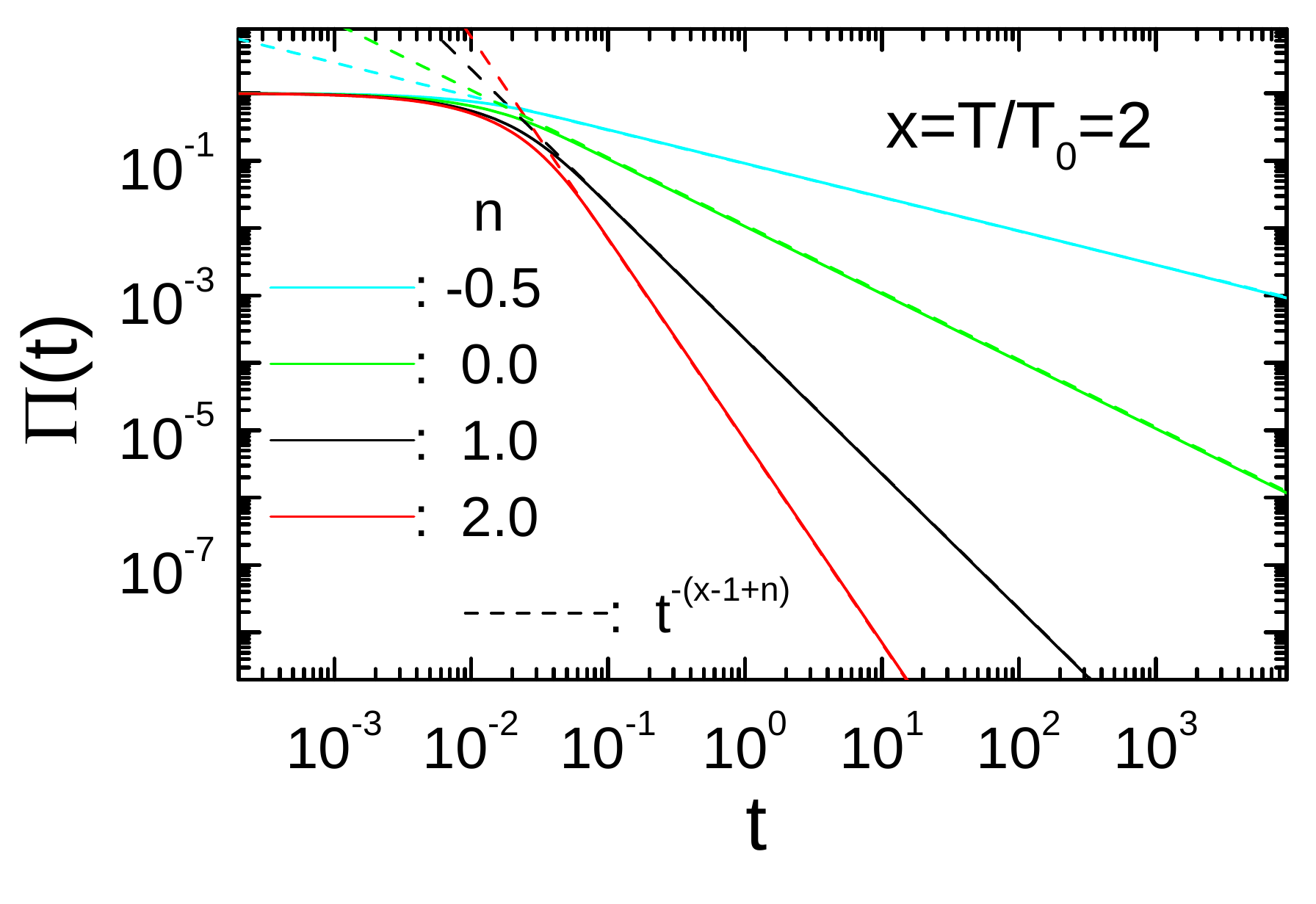}
\vspace{-0.5cm}
\includegraphics[width=8.0cm]{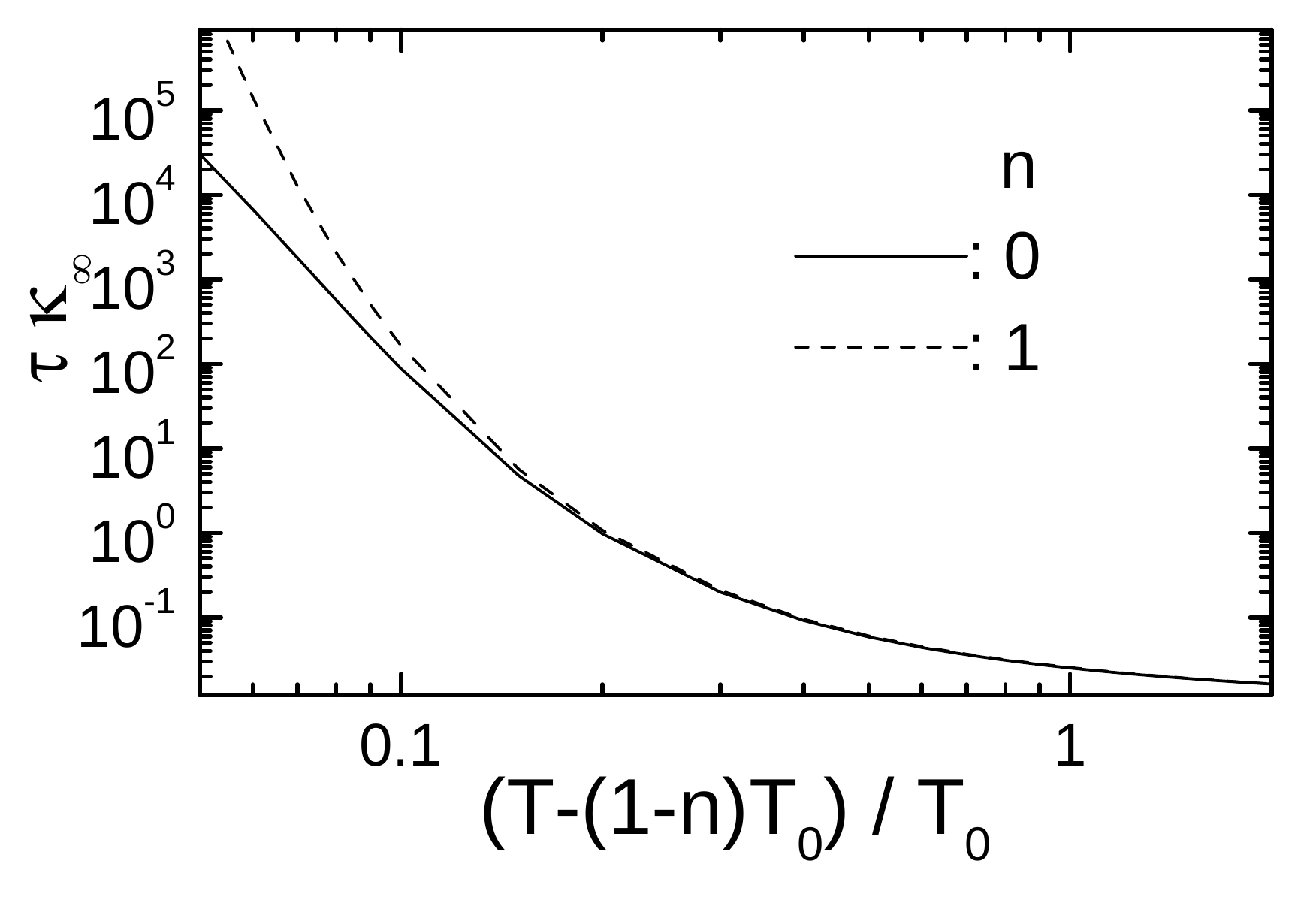}
\vspace{-0.0cm}
\caption{{\bf a)} (left panel): Two-time correlation function $\Pi(t)$ for $T=2T_0$ and various values for the $n$. 
The dashed lines correspond to the asymptotic behavior.
{\bf b)} (right panel): relaxation time $\t$ (defined via $\Pi(\t)=1/e$) versus reduced temperature for $n=0$ and $n=1$.
}
\label{Figur1}
\end{figure}

The linear susceptibility, which is the fourier transform of $\Pi(t)$, is given by 
\be\label{Chi1}
\chi_1(\om)=\b\int_0^\infty\!d\e\lg M(\e)^2\rg p^{\rm eq}(\e){\k(\e)\over\k(\e)-i\om}
\ee
and the static susceptibility follows from this to be 
\be\label{Del.chi1}
\D\chi_1=\chi_1(0)=\b{x-1\over x-1+n}
\ee
which diverges at a temperature $T=(1-n)T_0$.
Furthermore, the integral relaxation time
\be\label{tau.eq}
\tau_{eq}^{(n)}=\int_0^\infty\!dt\Pi(t)={1\over\k_\infty}{x-1\over x-2+n}
\ee
diverges at $T=(2-n)T_0$ which reduces to the well-known result $2T_0$ for $n=0$\cite{MB96}.
In Fig.\ref{Figur1}a), $\Pi(t)$ is shown for various values of $n$.
It is obvious that the asymptotic behavior is given by $\Pi(t)\sim t^{-(x-1+n)}$.
This means that for the model with an exponential DOS the choice of the dynamical variables used here has a strong impact on the temperature dependent dynamical properties as has been discussed before\cite{FS02}.
In addition, the relaxation time $\t$ that is defined as the time after which $\Pi(t)$ has decayed to $1/e$ shows a divergence at 
$T=(1-n)T_0$, cf. Fig.\ref{Figur1}b).
For $n=0$, this is a well known feature of the model\cite{MB96}.

One prominent difference between the model with an exponential DOS and the one with a Gaussian DOS is that the choice (\ref{M2e.exp}) does strongly change the shape of $\chi_1(\om)$ in the first case, while for the Gaussian model this shape is unaltered for different $n$ apart from a frequency-rescaling.
Only the amplitude strongly depends on the choice of $n$.
\section*{III. Nonlinear response functions}
In most of the experimental and theoretical treatments of the nonlinear response of supercooled liquids the modulus of the response functions
\be\label{X.alfa.def}
X_\a(\om,T)={T\over(\D\chi_1)^2}|\chi_3^{(\a)}(\om)|\quad,\quad\a=1,\;3
\ee
has been considered. 
Here, $\a=1$ denotes the one-$\om$ and $\a=3$ the three-$\om$ component of the modulus of $\chi_3(\om)$ and $\D\chi_1$ is the static linear susceptibility.

The expressions for the frequency-dependent response functions are computed using the general expressions obtained from time-dependent perturbation theory as shown in I\cite{G75}.
In general, the cubic response function reads
\be\label{Chi.a.3}
\chi^{(3)}(t)
=
{H_0^3\over2}\left[e^{-i\om t}\chi_3^{(1)}(\om)+e^{-i3\om t}\chi_3^{(3)}(\om)+c.c.\right]
\ee
where $c.c.$ denotes the complex conjugate and it has been assumed that the initial transients are died out.
In I, also all relevant expressions for the trap model are given and the results for the Gaussian model are discussed.
It has been found that the response depends on the details of the coupling of the external field to the transition rates and in particular on the coupling strength to the initial or the final state of a transition. 
More important, however, is the dependence of the cubic susceptibility on the dynamical variable chosen, i.e. on the value of $n$ in 
eq.(\ref{M2e.exp}).
Because for the cubic response one has to deal with averages of products of four realizations of the dynamical variable $M$, a simple Gaussian approximation for such four point functions has been used in I and II and will also be employed in the present paper (along with eq.(\ref{M2e.exp})):
\Be\label{Mh4.mit.Gauss}
\lg M(\e_1)M(\e_2)M(\e_3)M(\e_4)\rg
&&\hspace{-0.6cm}=
\d(\e_1-\e_2)\d(\e_3-\e_4)e^{-n(\e_1+\e_3)}
\nonumber\\
&&\hspace{-0.6cm}+\;
\d(\e_1-\e_3)\d(\e_2-\e_4)e^{-n(\e_1+\e_2)}
\\
&&\hspace{-0.6cm}+\;
\d(\e_1-\e_4)\d(\e_2-\e_3)e^{-n(\e_1+\e_2)}
\nonumber
\Ee
Note that this approximation cannot be justified rigorously. However, it appears very meaningful in the context of a mean-field model like the one considered here.
In the following discussion, I will present results obtained for a coupling of the field to the initial state of a transition, i.e. 
$\mu=1$ and $\g=0$ in the terminology of I.
This is sufficient to discuss the general behavior of $\chi_3^{(\a)}(\om)$ because other choices of $\mu$ and $\g$ give rise to small quantitative changes only.

\begin{figure}[h!]
\centering
\includegraphics[width=8.0cm]{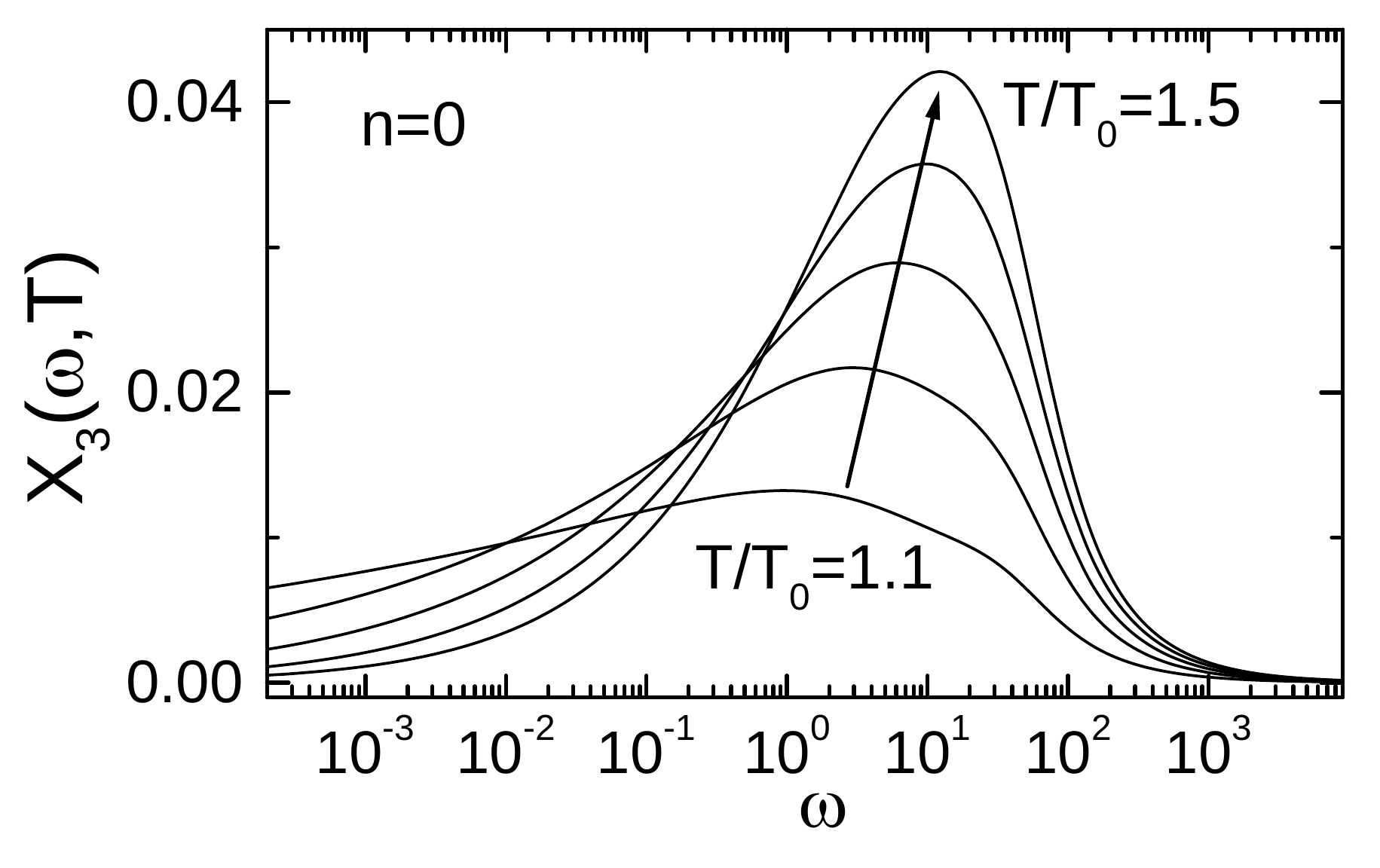}
\vspace{-0.5cm}
\includegraphics[width=7.9cm]{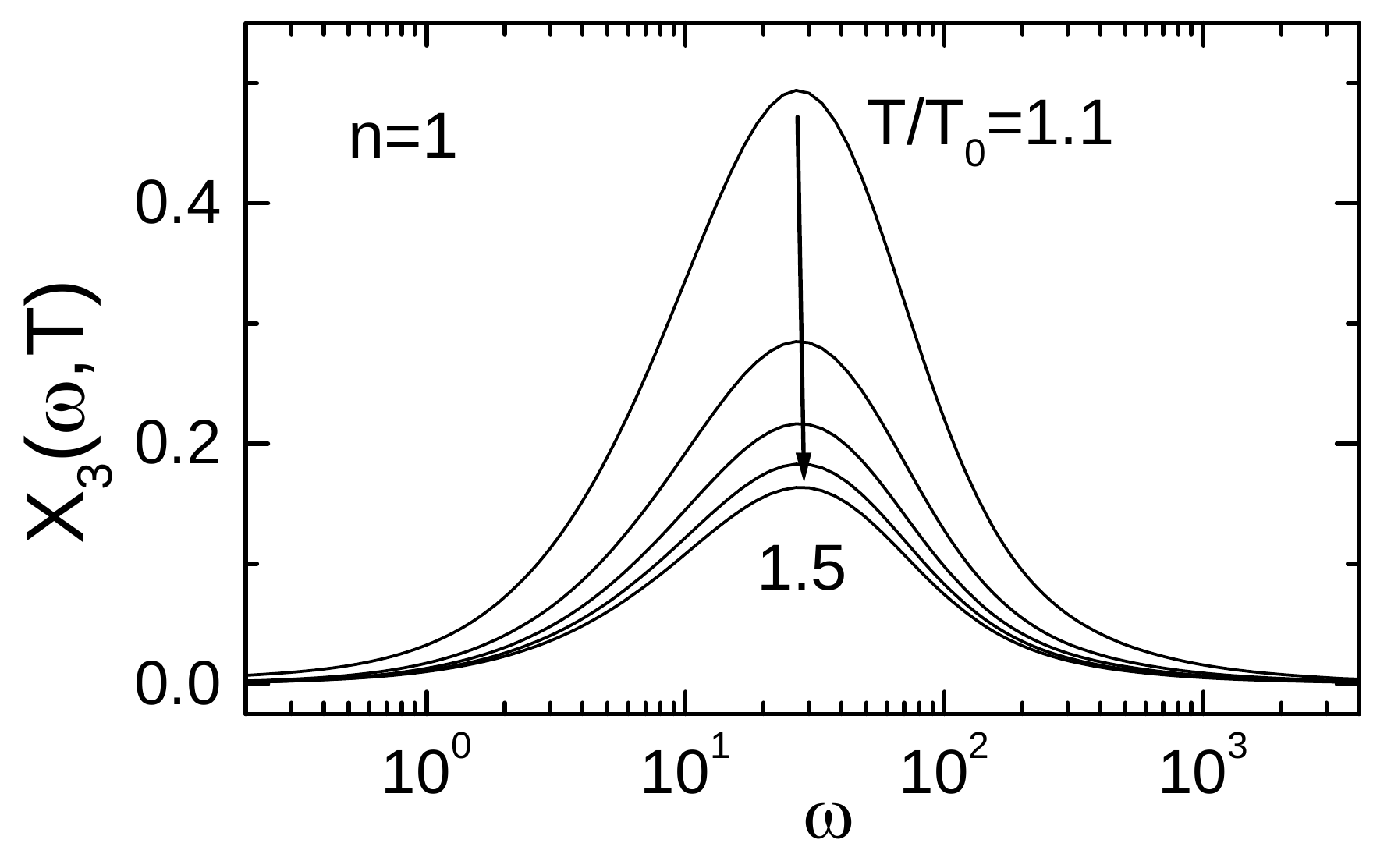}
\vspace{-0.0cm}
\includegraphics[width=7.9cm]{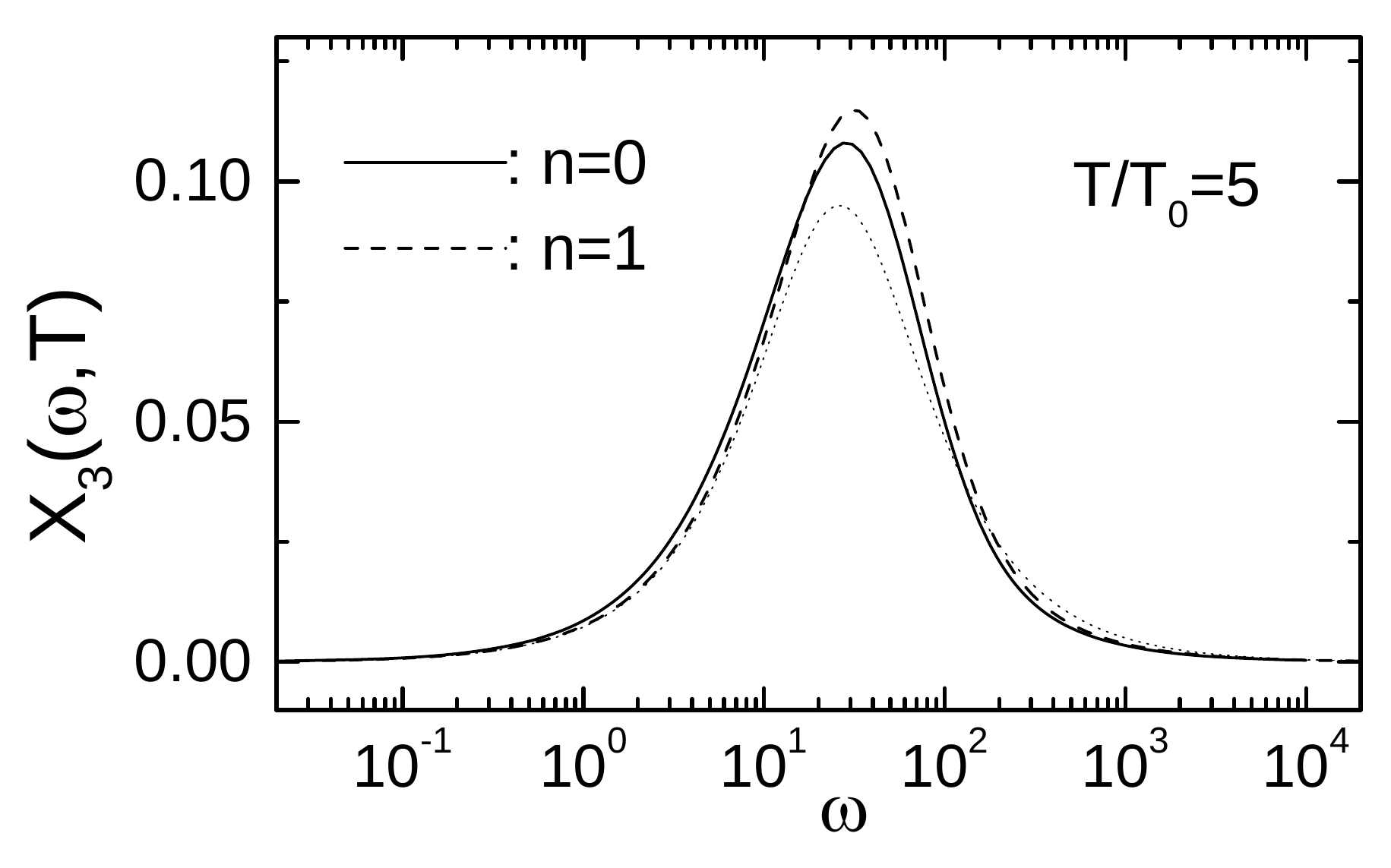}
\vspace{-0.0cm}
\caption{$X_3(\om,T)$ versus frequency for $\mu=0$ and various temperatures ($T/T_0=1.1$, $1.2$, $1.3$, $1.4$, $1.5$).
{\bf a)} (upper left panel): $n=0$, the dynamical variable are independent of the trap energy.
{\bf b)} (upper right panel): $n=1$, the dynamical variable depend on the trap energy in an Arrhenius-like way.
{\bf c)} (lower panel): $X_3(\om,T)$ for $T/T_0=5$ and $n=0$, $n=1$. The thin dotted line is a Lorentzian for comparison.
}
\label{Figur2}
\end{figure}

In Fig.\ref{Figur2}a) $X_3(\om,T)$ is shown for variables that are independent of the trap energy, $n=0$.
It is obvious that $X_3$ exhibits a peak with a maximum value that grows with increasing temperature.
Furthermore, the shape of $X_3(\om,T)$ becomes more symmetric on the logarithmic scale with increasing temperature.
Fig.\ref{Figur2}b) shows  $X_3(\om,T)$ for an energy dependent variable with $n=1$.
Also in this case a peak is observed the height of which, however, decreases with increasing temperature.
The behavior for these two values of the parameter $n$ is qualitatively the same as found for the Gaussian trap model in I\cite{G75}.
(It has to be noted that for comparison one has to replace $n$ by $(-n)$ in paper I due to the different definition of the escape rates.)
Also for other values of $n$ and also for the one-$\om$ component $X_1(\om)$ the qualitative features are similar for the exponential trap model considered here and the Gaussian one.
In particular, for negative values of $n$ one finds a trivial behavior at low temperatures and the occurence of a peak at higher temperature.
It is obvious from Fig.\ref{Figur2}c) that for high temperatures $X_3(\om)$ becomes independent of the choice of the dynamical variable and exhibits a symmetric shape.
This fact can be understood from the fact that $e^{-n\b\e}$ becomes independent of $n$ for vanishing $\b$.

An important quantity is the low-frequency limit $X_\a(0)$ because the value of $X_\a(0)$ relative the maximum value of 
$\chi_3^{(\a)}"(\om)$ determines the gross behavior of the modulus.
In particular, if $X_\a(0)$ is small a peak might be observed while for large $X_\a(0)$ usually trivial behavior is to be expected.
This quantity is found to be given by:
\be\label{X3.Null}
X_3(0)={n(n-1)\over8}{x-(1-n)\over(x-1)(x+n)(x-(1-2n))}
\quad;\quad X_1(0)=3X_3(0)
\ee
which vanishes for $n=0$, $1$ and $n=1-x$.
Additionally, it diverges for $T=T_0$, $T=-nT_0$ and $T=(1-2n)T_0$.
For the Gaussian model one has $X_3(0)=(1/8)e^{n(n-1)\b^2\s^2}|e^{n\b^2\s^2}-1|$ which also vanishes for $n=0$, $1$ but of course does not diverge at any finite temperature.
The particular structure of the expression for $X_\a(0)$ can be understood from the quantities involved in the calculation, cf. I.
One has $\chi_3^{(\a)}\propto(\xi_2-\xi_1)$ with $\xi_1$ and $\xi_2$ determined by averages over the dynamical variable $M$.
In particular, $\xi_1=(x/(x+n))(x-1)/(x-1+n)$ is related to the static linear susceptibility and the divergence at $T=(1-n)T_0$ is compensated by the denominator in the definition of $X_\a(0)$, eq.(\ref{X.alfa.def}).
Furthermore, $\xi_2=(x-1)/(x-(1-2n))$ depends on the square of the second moment of the distribution of $M$, 
$\xi_2=\overline{\lg M^2\rg^2}_T$.
The form of $\xi_2$ can be understood from the Gaussian approximation for the averages of $M$, which effectively results in a replacement of $\overline{\lg M^4\rg}_T$ by $\overline{\lg M^2\rg^2}_T$ and is responsible for the occurence of the divergence at $T=(1-2n)T_0$.
Eq.(\ref{X3.Null}) shows that for negative values of $n$ this additional divergency shifts the accessible temperature range to higher temperatures.

\begin{figure}[h!]
\centering
\includegraphics[width=8.0cm]{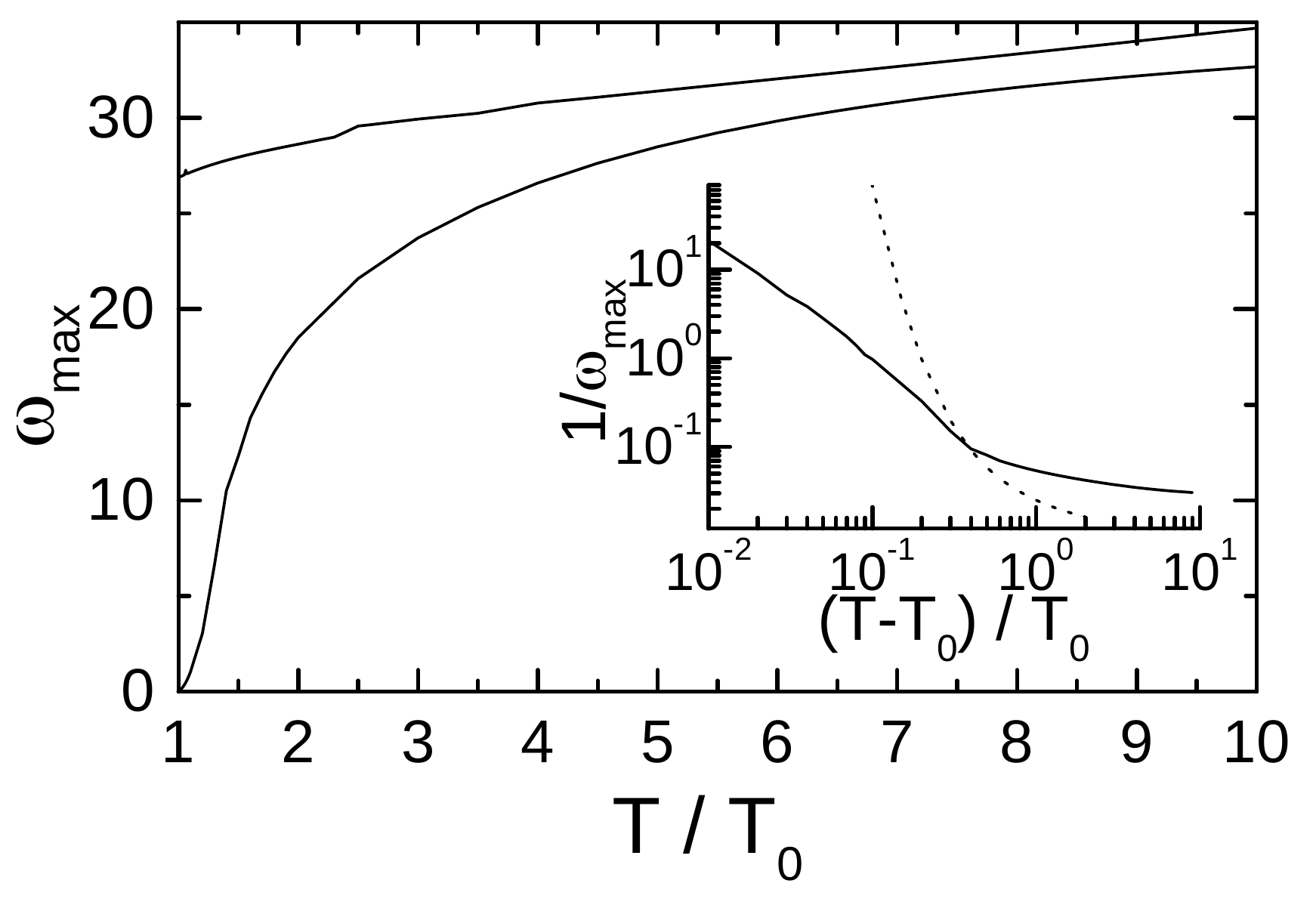}
\vspace{-0.5cm}
\includegraphics[width=8.3cm]{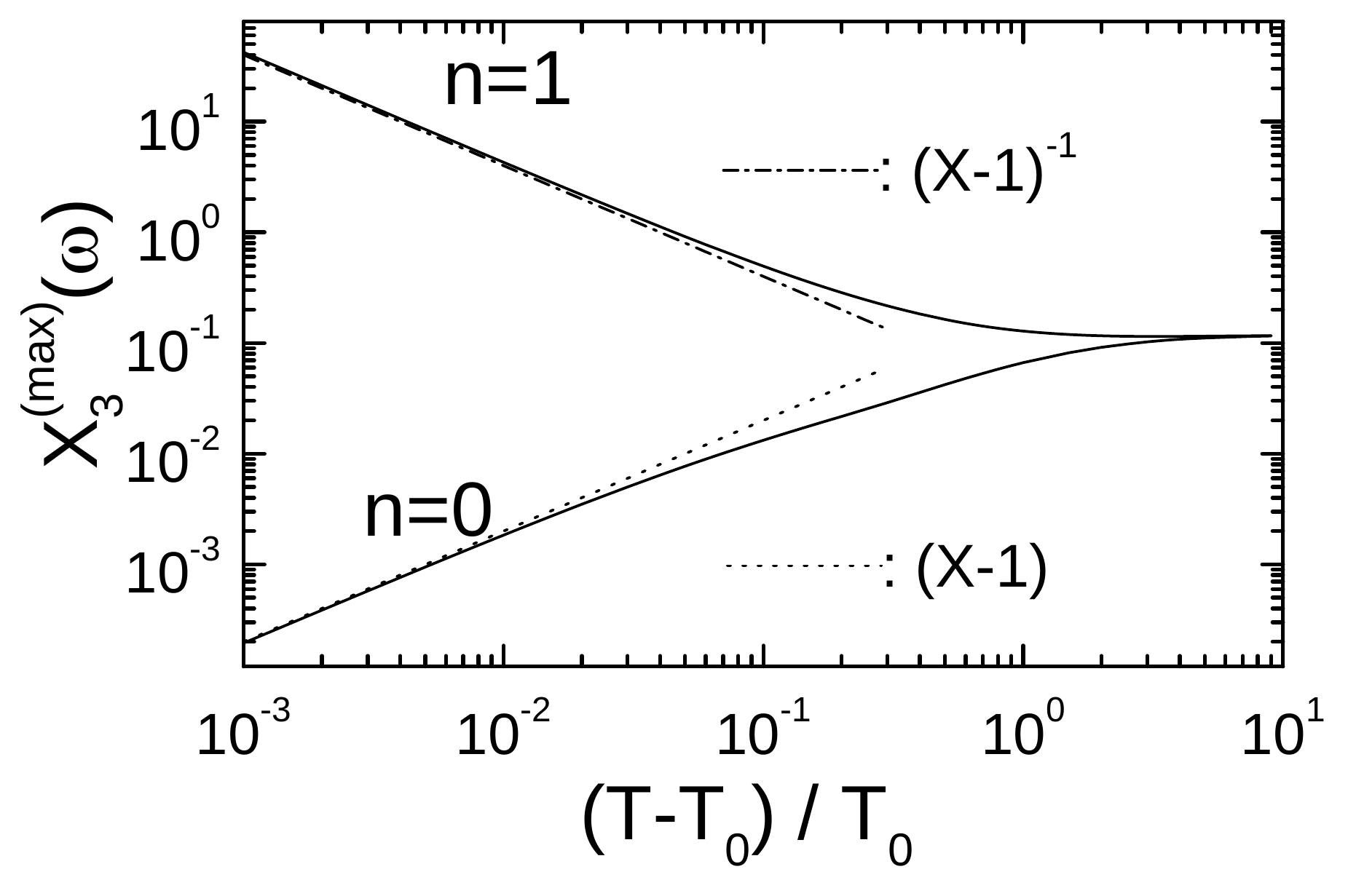}
\vspace{-0.0cm}
\caption{
{\bf a)} (left panel): Maximum frequency of $X_3(\om,T)$, $\om_{\rm max}$, as a function of temperature for $n=0$ (lower curve) and $n=1$ (upper curve).
The inset shows the characteristic time scale $1/\om_{\rm max}$ for $n=0$ versus reduced temperature. For comparison the relaxation time 
$\t$ is shown as the dotted curve, cf. Fig.\ref{Figur1}b).
{\bf b)} (right panel): Maximum of $X_3$, $X_\a^{(\rm max)}=X_\a(\om_{\rm max})$ versus reduced temperature, $(T-T_0)/T_0$ for $n=0$ (lower curve) and $n=1$ (upper curve).
}
\label{Figur3}
\end{figure}

In Fig.\ref{Figur3}a) the peak frequency $\om_{\rm max}$ of $X_3(\om,T)$ is plotted as a function of temperature for $n=0$ and $n=1$.
While for $n=1$ $\om_{\rm max}$ only shows a weak temperature dependence, this is different for $n=0$.
The inset shows the inverse peak frequency as a function of reduced temperature for $n=0$.
While the temperature dependence is not as strong as the corresponding one of the relaxation time $\t$, it is still appears to show some scaling behavior for $T\to T_0$.
However, as can be observed already from $X_3(\om)$ shown in Fig.\ref{Figur2}a) the spectra become very broad for low temperatures and the determination of $\om_{\rm max}$ becomes increasingly difficult.
It should be noted furthermore, that the mild temperature dependence of $\om_{\rm max}$ for $n=1$ for $T>T_0$ can be understood from the fact that the corresponding glass transition takes place at $T=0$, cf. Fig.\ref{Figur1}b).

In Fig.\ref{Figur3}b), the maximum value of $X_3$, $X_3^{(\rm max)}=X_3(\om_{\rm max})$, is plotted versus reduced temperature for $n=0$ and $n=1$.
This value is easier to determine also for broad spectra.
For $T\to T_0$, $X_3^{(\rm max)}$ behaves as $(T-T_0)^{-1}$ $(n=1)$ or as $(T-T_0)$ $(n=0)$. 
With increasing temperature the nature of the dynamical variable (the value of $n$) becomes irrelevant and the different response functions behave very similar, cf. the discussion in the context of Fig.\ref{Figur2}c).

Additionally, in II\cite{G76} the approximation introduced by Bouchaud and Biroli\cite{Bouchaud05} has been discussed for the Gaussian trap model.
The most important contribution to the response near a phase transition has been shown to be given by
\be\label{R3.FDT}
R_3(t_0,t_1,t_2,t_3)\sim\b^3{d^3\over dt_1dt_2dt_3}\lg M(t)M(t_1)M(t_2)M(t_3)\rg
\ee
Here, $\lg M(t)M(t_1)M(t_2)M(t_3)\rg$ ($t_0>t_1>t_2>t_3$) is the four-time (4t) analogue of the two-time correlation function 
$\lg M(t_1)M(t_0)\rg$, cf. eq.(\ref{C2.Pi}), and $R_3(t_0,t_1,t_2,t_3)$ denotes the (impulse) response.
Note that this relation has the very appealing form of a quasi fluctuation dissipation theorem (FDT).
As mentioned in the Introduction, it was found in II that this relation unfortunately does not give a sound description of the cubic response for the Gaussian trap model due to a wrong estimate of the low-frequency behavior.

For the exponential trap model, eq.(\ref{R3.FDT}) also does not work, because one can show analytically that 
\be\label{Chi3.FDT.0}
X_{3;FDT}(0)={1\over4}\left|{(x-1)(x+n)\over x^2(x-1+n)(x-2)}\right|
\ee
and $\chi_{3;FDT}^{(1)}(0)=3\chi_{3;FDT}^{(3)}(0)$.
It is evident that $\chi_{3;FDT}^{(3)}(0)$ exhibits a divergence at $T=2T_0$ that is absent in $\chi_{3}^{(3)}(0)$.
This divergence has its origin in the dependence of $X_{3,FDT}(0)$ on the relaxation time 
$\t_{eq}^{(0)}=\k_\infty^{-1}(x-1)/(x-2)$.
It is obvious that at variance with the exact expression, eq.(\ref{X3.Null}), $X_{3,FDT}(0)$ does not vanish for $n=0$ or $n=1$.
If one considers the full modulus $X_{3,FDT}(\om)$, it is found that one has trivial behavior for a large variety of values for $n$, i.e. for different dynamical variables.
Thus, also for the exponential trap model the FDT-like approximation for the cubic response function does not represent the exact expression properly.
The main reason for the failure of eq.(\ref{R3.FDT}) has its origin in the fact that the temporal change of the 4t-correlation function takes place on the diverging time scale of the 2t-correlation function, which does not hold for the full response, 
cf. Fig.\ref{Figur3}a).
\section*{IV. Conclusions}
The nonlinear response for the trap model with an exponential DOS shows a behavior that strongly depends on the dynamical variable chosen.
In contrast to the model with a Gaussian DOS, the exponential trap model exhibits a glass transition at $T_0$ that is reflected in all dynamical observables.
However, if one considers temperatures that are not close to $T_0$ the two models behave very similar, a fact that has been noticed earlier, see e.g.\cite{G73}.
Here, I have restricted the focus on the high temperature phase, where the system always reaches equilibrium.
Generally, the behavior of the cubic response in the form of the modulus depends strongly on the low-frequency limit $X_\a(0)$ because the relative height of $\chi_3^{(\a)}(0)$ and $\chi_3^{(\a)}"(\om_{\rm max})$ determines whether $X_\a(\om)$ exhibits a peak or shows a 'trivial' decay from the non-vanishing $X_\a(0)$ to $X_\a(\infty)=0$.
For special choices of the dynamical variable ($n=0$, $1$), $X_\a(0)$ vanishes for all temperatures and therefore the divergences occuring for other values of $n$ are absent.
The divergence at $T=T_0$ is generic for the trap model and the one at $T=-nT_0$ has ist origin in the form chosen for the dynamical variables, eq.(\ref{M2e.exp}).
The additional divergence at $T=(1-2n)T_0$ stems from the Gaussian factorization approximation for the four-point functions, cf. eq.(\ref{Mh4.mit.Gauss}) and this might change for other approximations for the four-point correlation.
However, in view of the mean-field nature of the fully connected trap model, the Gaussian approximation appears meaningful and consistent internally.

A prominent difference between the exponential and the Gaussian trap model, apart from the absence of a glass transition in the latter, is the importance of the nature of the dynamical variable $M(t)$ as chosen in eq.(\ref{M2e.exp}) on the linear response is much stronger in the exponential model.
While $\chi_1(\om)$ in case of the Gaussian model does not change in a scaled representation, cf. the discussion in 
paper I\cite{G75} this is completely different for the exponential model.
Here, also the frequency dependence is strongly affected by the choice of $n$, cf. the behavior of $\Pi(t)\sim t^{-(x-1+n)}$.
This directly translates to an explicit $n$-dependence of $\chi_1(\om)$.
In addition, the maximum value of the cubic response shows an interesting scaling behavior for $n=0$ and $n=1$, Fig.\ref{Figur3}b) while for the Gaussian model there is hardly any temperature dependence for $n=0$.

In the trap model, the relevant time scale is the temperature dependent relaxation time of the two-time correlation function and  the frequency dependence of the cubic response basically is similar to the corresponding one of the linear susceptibility.
For the exponential trap model, the value of $X_3(\om_{\rm max})$ shows a scaling behavior as a function of reduced temperature for special choices of the dynamical variable ($n=0$, $1$). 
Note that in the trap model the height of the peak has no relation to a number of correlated particles 
$N_{\rm corr}$\cite{CrausteThibierge10, Brun11} due to the mean-field nature of the model.
Of course, the situation in glassforming systems is vastly different and the length scale of this is the last candidate.  next esc will revert to uncompleted text. he heterogeneities might well determine the peak height.
However, the humped shape cannot be taken as a unique feature of 'glassy correlations'.

Another finding of the present calculation is that the quasi-FDT approximation, eq.(\ref{R3.FDT}), does not give a good description of the nonlinear response for the trap model.
Due to a wrong estimate of $\chi_{3,FDT}^{(\a)}(0)$ one has only trivial behavior in this approximation and no peak is observed.
In particular, the critical behavior of $\chi_3^{(\a)}(\om)$ and $\chi_{3,FDT}^{(\a)}(\om)$ are different because the respective low-frequency limits diverge at different temperatures.

The exponential trap model shows a glass transition albeit without any length scale.
Therefore, it would be interesting to study models with a diverging length scale and compare the full cubic response function to the quasi-FDT approximation, in particular in the vicinity of the critical point.
Furthermore, the nonlinear response in the glassy phase of the exponential trap model where aging is important will be investigated in a future work.

In conclusion, I have computed the nonlinear response for an exponential trap model in the high temperature phase and found a behavior that is similar to that of the Gaussian trap model if temperatures that are not too close to the glass transition are considered.
The modulus shows a behavior that strongly depends on the choice of the dynamical variable that is used to monitor the response and one finds a peak or trivial behavior depending on this choice.
\section*{Acknowledgement}
I thank Roland B\"ohmer, Jeppe Dyre and Gerald Hinze for fruitful discussions on the topic of this paper.
%
%

%
\end{document}